\documentclass{ws-ijmpd}

\newcommand{\beq}{\begin{equation}}
\newcommand{\eeq}{\end{equation}}
\newcommand{\bey}{\begin{eqnarray}}
\newcommand{\eey}{\end{eqnarray}}

\newcommand{\kpc}{\, {\rm kpc} }

\newcommand{\grad}{{\bf \nabla}}

\newcommand{\be}{\begin{equation}}
\newcommand{\ee}{\end{equation}}
\newcommand{\ba}{\begin{eqnarray}}
\newcommand{\ea}{\end{eqnarray}}

\newcommand{\kms}{\, {\rm km}s^{-1}}

\newcommand{\thetab}{\mbox{\boldmath $\theta$}}

\newcommand{\tg}{{g}}

\newcommand{\R}{\mbox{\boldmath $R$}}

\begin{document}

\markboth{HongSheng Zhao}
{Lensing by an Uneven Dark Energy}

%
\catchline{}{}{}{}{}
%

\title{ An Uneven Vacuum Energy Fluid as  
$\Lambda$, Dark Matter, MOND and Lens }

\author{HongSheng Zhao}

\address{
National Astronomical Observatories, Chinese Academy of Sciences, 
\\Datun Road, Chaoyang district,  Beijing, China \\
SUPA, School of Physics and Astronomy, University of St Andrews\\
KY16 9SS, Fife, UK\\
hz4@st-and.ac.uk}

\maketitle

\begin{history}
\received{Based on lectures at the Lensing Winter School, Sicily 2006 \& 
invited talk at Fundamental Physics From Quantum to Cosmology, Washington DC 2006}
\revised{3 Mar 2008}
\comby{Managing Editor}
\end{history}

\begin{abstract}
Various TeVeS-inspired and f(R)-inspired theories of gravity 
have added an interesting twist 
to the search for dark matter and vacuum energy, modifying the landscape of
astrophysics day by day.  These theories can be together called 
a {\bf N}on-{\bf u}niform Dark Energy fluid 
(a Nu-Lambda fluid or a ${\mathbf V\Lambda}$ fluid); 
a common thread of these theories, according of an up-to-date summary by HZL \cite{Halle}, is a non-uniform vector field, describing an uneven vacuum energy fluid.  The so-called "alternative" gravity theories are 
in fact in the standard GR gravity framework except that 
the cosmological "constant" is replaced by a non-trivial non-uniform vacuum energy,   
which couples the effects of Dark Matter and Dark Energy together by a single field.
Built initially bottom-up rather than top-down as most gravity theories, 
TeVeS-inspired theories are healthily rooted on empirical facts.  
Here I attempt a review of some sanity checks of these fast-developing theories 
from galaxy rotation curves, gravitational lensing and cosmic acceleration.  
I will also discuss some theoretical aspects of the vacuum energy,
and point out some analogies with electromagnetism and the Casimir effect.  
\end{abstract}



\section{The three pillars of the standard $\Lambda$CDM cosmology}

The standard cosmological paradigm is built on three pillars: 
Einsteinian gravity, a cosmological constant or vacuum energy density 
about $10^{-10}$ erg/cm$^3$ due to unknown physics, 
and a thermal relic of Cold Dark Matter due to physics at the TeV scale.
While the independent experimental basis of each of the three is debatable on
astronomical scales, but their synergy (characterised by the
cosmological pie diagram) has proven amazingly successful at describing the
Universe especially on large scale.

Despite its apparently enticing simplicity, the paradigm leaves much to
be understood and is challenged by observations on
galaxy scale.  For example, the
experimentally undetected dark matter is generally thought to be
Minimal SuperSymmetry Model (MSSM) particles, 
and is predicted to be cold and clump in scale-free fashion,
while observations of dwarf galaxies suggest the particles are warm 
with a kpc-scale, below which 
DM is smoothed out by free-streaming of the thermal motion.  

Most embarassingly is that there is no physics for the cosmological constant;
The MSSM physics at TeV scale 
fails to explain the tiny vacuum energy of the universe by 120 orders of magnitude.   
This is regarded by many theoreticians as evidence for new physics at low energy scales.

\subsection{A characteristic scale for both Dark Matter and Dark Energy}

As an important puzzle about dark matter, it
has long been noted that on galaxy scales dark matter and baryonic
matter (stars plus gas) have a remarkable correlation, and respect a
mysterious acceleration scale $a_0 \sim 1$ Angstrom per second squared. 
\cite{Mil} \cite{BM84} \cite{SM02} \cite{SSM05}

The Newtonian gravity of the known matter (baryons, neutrinos, eletrons, etc.) 
${\bf g}_{K}$  and the dark matter gravity
${\bf g}_{DM}$ are correlated through an empirical relation
\cite{FGBZ} \cite{VL} such that the light-to-dark ratio, 
experimentally determined to fit rotation curves, satisfies
a very simple relation
\beq\label{gdm}
g_{DM} \approx \sqrt{g_{K} a_0}, \qquad a_0 \equiv 1 {\rm Angstrom}\,{\rm sec}^{-2}
\eeq
where $a_0$ is the fore-mentioned gravity scale, below which DM and DE 
phenomena start to surface.  This DM-to-baryon relation fits rotation curves of 
faint and bright spiral galaxies fairly well (cf. Fig.~\ref{vclens}).

Such a tight correlation is difficult to understand in a galaxy
formation theory where dark matter and baryons interactions enjoy 
huge degrees of freedom.  This spiral galaxy based empirical relation
is also consistent with some elliptical galaxies and gravitational
lenses.

It is also hard to explain from fundamental physics why vacuum energy
starts to dominate the Universe density only
at the present epoch, hence marking the present as the turning point for 
the universe from de-acceleration to acceleration.

The puzzles of DM and DE are related by the fact that 
\beq
a_0 \sim \sqrt{\Lambda} \sim c H_0.
\eeq 
{\it Somehow dark energy and dark matter are tuned to shift dominance when
the energy density falls below ${a_0^2 \over 8 \pi G} \sim 10^{-10}$erg/cm$^3$}.  
These empirical facts should not be completely treated as random
coincidences of the fundamental parameters of the universe.
The explanation with standard paradigm has been unsatisfactory.

The problems of $\Lambda$CDM have led some to believe the paradigm is an
effective theory, e.g., a 4D projection of a more fundamental 5D brane
world theory.  Some also question the Einsteinian gravity since its
associated equivalence principles, remain untested on galaxy scale and
cosmological scale.  
A less drastic approach is to keep the framework of the Einsteinian 
gravity, but design the Lagrangian  
for the dark energy field to have the effect of dark matter as well.
An example of the latter approach is the Vector-for-$Lambda$ model or the
${\bf V}\Lambda$ model of Zhao\cite{VL}, where a photon-like but massive 
vector field is speculated to exist even in vacuum.  A careful choice of 
the dark energy field can replace the role of dark matter too, i.e., 
the DM and DE parts of the cosmic pie diagram are in fact 
two aspects of a single species of dark fluid.

\section{Energy Density of the Uneven Vacuum}

A common way to probe dark matter in galaxies is 
gravitational lensing.  The amount of light bending is 
an indicator of the non-flatness of the space-time metric, hence 
constraining the matter distribution.
However, light bending is a general property of propagation of E\&M
waves following Fermat's principle, or the geodesics.  The amount of bending can be 
an indicator of the non-uniformness of the propagated medium, 
e.g., in the case of atmospherical seeing.
Light could be bent even in the vacuum because 
the vacuum is not empty, and can be a fluid of certain energy density.   

The energy density in the vacuum can vary 
with space and time as well.  
It is interesting that the Casimir effect predicts in principle a pressure 
${\hbar c \pi^2 \over 240 \Delta^4} \sim 10^{-10}$ erg/cm$^3$ 
for two neutral metal plates separated by a distance $\Delta \sim 0.01$cm.
This pressure can drive the plates closer and closer, 
because the zero-point of the vacuum energy density due to eletromagnetic
waves between the plates is lower than outside the plates; as the plates close in 
the pressure goes up as $\Delta^{-4}$.    
The Casimir effect is indeed observed experimentally when the plates are separated by  $\Delta=100$ nanometer or closer.  The vacuum energy density could fluctuate 
spatially, too.\footnote{E.g., the vacuum energy 
due to eletromagnetism would not be uniform  
if many Casimir plates were randomly distributed, or 
if these Casimir plates were replaced by a distribution of polarisable neutral atoms in the universe.  An analogous situation 
(although with a different physics from the Casimir effect) 
happens in solid-state physics, where 
the effective dielectric "constant" $\epsilon$ can be spatially varying.  
As an effect, e.g., the normally $r^{-2}$ repulsive force between two electrons 
becomes a complicated function of their separation 
if they are inside a lattice of polarisable neutral atoms,
and can even change the sign in special cases 
\cite{Friedberg}.}

Likewise, for very different physics, the zero-point of the vacuum 
could fluctuate spatially or evolve time-wise due to gravitational physics.
E.g., the universal vacuum energy density during inflation is much higher than
the vacuum energy density today.  Any spatial variation of the vacuum 
energy density would generate more curvature 
in some patches of space-time, creating a dark-matter-like effect.  
The vacuum in this case appears as a dark fluid with fluctuations.  
The effects of fluctuation might manifest as a temporal or spatial change of 
the gravitational coupling factor\cite{Odintsov} $G_{\rm eff} = G/\mu(t,{\bf r})$, 
where $G$ is the usual gravitational constant determined in earth-based labs,
and $\mu$ is some kind of dielectric-like parameter, which can determined 
in a {\it Gedanken} experiment by 
$ {G \over \mu} \equiv { \left|\ddot{\bf r}_1-\ddot{\bf r}_2 \right| \over (m_1+m_2) \left|{\bf r}_1-{\bf r}_2\right|^{-2} }, $
where one 
measures the relative acceleration $\left|\ddot{\bf r}_1-\ddot{\bf r}_2 \right| $ of two neutral test particles of $m_1$ and $m_2$ 
slightly separated by a distance $\left|{\bf r}_1-{\bf r}_2\right|$
in a table-top Cavendish-type experiment near the space-time coordinate $(t,{\bf r})$
in the intergalactic space.  

The gravity at a typical place in a galaxy is very weak, 
is about a factor $4 \times 10^4$ smaller than the solar gravity on Pluto.
E.g., the Sun's acceleration around the Galaxy 
\beq 
g \sim {(200 \kms)^2 \over 10 \kpc} 
  \sim {{\rm LightSpeed} \over 10 \times {\rm HubbleTime}}
  \sim {1 {\rm m} \over {\rm day}^2} 
  \sim {1 {\rm Angstrom} \over {\rm sec}^2} .
\eeq 

The gravitational energy density associated with 
$1$Angstrom per second squared gravitational field is
about $10^{-10}$ erg/cm$^3$.  This is roughly the scale of 
the cosmological constant, yet 
$10^{-12}$ smaller than the current experimental sensitivity in the Casimir pressure.
New physics on such weak scale is allowed as far as experiments are concerned.

A mundane example of $1$Angstrom per second squared gravity is 
{\it the mutual Newtonian 
gravity of two nearly parallel sheets of printing papers} approximately.
The gravitational attraction of two sheets of paper could depend
on environment.  Consider a {\it Gedanken} experiment with 
a gravitationally torquing pendulum made by two 
misaligned suspended sheets of paper.  If one could measure
the period of the torquing pendulum not only here on Earth 
(as in free-fall experiments in an Einstein tower), 
but also take the table-top experiments to 
the edge of the solar system (where Pioneer 10/11 probes are),
in the interstellar space (where galactic stars orbit)
and in the expanding void between galaxies, then 
one could measure how $G_{eff}$ changes with space and time.

\section{TeVeS-like modified gravity: motivations and challenges}

Modifying gravity is a recurring exercise which started ever since
the general acceptance of Einsteinian gravity, which was itself a
revolutionary modification to Newtonian gravity.  Many theories modify
the Einstein-Hilbert action to introduce a new scalar field which
manifests itself only through the extra bending of space time, but its
coupling to the metric is different from the simple coupling of
massive particles with the space-time metric.

By construction, the theories would respect Special Relativity
prescription of metric co-variance, and preserve conservations of
momentum and energy.  They do allow for a table-top Cavendish-type
experiment with a torquing pendulum to measure an effective
gravitational constant $G_{eff}(t,x)$ which varies with time and
environment of the experiment.  For example, the recent $F(R)$ models
are motivated to replace the cosmological constant with a vacuum energy density  
depending on the curvature of space-time, hence
evolving with the cosmic time in a way to drive the acceleration
of the universe at late time.

However, among two dozen theories proposed after GR, very few survive
the precise tests on SEP in the solar system and the well-studied
binary pulsars.  Even fewer are motivated and succeeded in addressing
both astronomical dark matter and cosmological constant.

Bekenstein's TeVeS\cite{Bek} 
is a first effort in the direction of solving outstanding problems.
Its partial success has spurred several variations of the theory, including 
Sanders' Bi-Scalar-Tensor-Vector theory \cite{BSTV}, Zlosnik et al.'s generalized 
Einstein-Aether theory \cite{ZFSfk}, and Zhao's Vector-for-$\Lambda$ model\cite{VL}.  These hold
the promise of explaining both dark matter and cosmological constant
by relaxing the SEP (strong equivalence principle) 
only in untested weak gravity environments like in galaxies, but 
respecting the SEP to high accuracy in the solar system.

Crudely speaking, such theories have an aether-like field with an
aquadratic kinetic term in its Lagrangian density, so the $G_{eff}$
can be made a function of the strength of gravitational energy 
${|g|^2 \over 8\pi G}$, such that
$G_{eff}$ is constant within $10^{-16}$ anywhere in the solar system,
yet {\it varies} by a factor of 10 in galaxies.  Enhancing the
$G_{eff}$ mimics the effects of adding dark matter.  The effects resemble dielectric.
E.g., in the $f(K_4)$ model of $V\Lambda$, the Poisson equation around 
a static galaxy of a baryonic density $\rho$ becomes \cite{Halle}
\beq \nabla \cdot \left[ {\bf E}- {\bf P} \right] =  \rho, \eeq
where ${\bf E}={ {\bf\nabla} \Phi \over 4 \pi G }$ is the rescaled 
gravity, 
remniscent of an electric field and 
${\bf P} = \lambda(|{{\bf E} \over \Pi_0}|) {\bf E} $ is a polarisation-like field with 
a susceptibility $\lambda$ being a function of the field strength $|{\bf E}|$
and a characteristic column density constant $\Pi_0$ comparable to that of 
a sheet of paper.   

\section{DM and DE as two faces of the same coin: Uneven Dark Energy fluid}

\subsection{Vector or scalar, modified or not?}

TeVeS-like theories, as GR, are single-metric theories. 
They can often be casted to the GR framework 
with a sophisticated Vacuum Energy term, all in physical metric \cite{Halle} \cite{ZFS}.

To see this, let $g_{\mu\nu}$ being the physical
metric, then near a quasi-static system like a galaxy, the physical space-time
is only slightly curved, and can be written as in terms of $x_0=ct$, and  
cartesian coordinates $(x_1,x_2,x_3)$ centred on the galaxy as
 \ba
    -c^2 d\tau^2 & \approx & - \exp\left({-2\Phi \over c^2}\right) dx_0^2 +  \exp\left({2\Phi \over c^2}\right) dl^2, \qquad dl^2=(dx_1^2+ dx_2^2+dx_3^2),
 \ea
where ${|\Phi| \over c^2} \ll 1$.  
To show that $\Phi$ takes the meaning of a gravitational
potential, we note that a non-relativistic massive particle 
moves along the geodesic equation (or Lagrangian equation)
 \beq
    \sum_{\beta=0}^3 {d^2 (g^{\alpha\beta} x_\beta) \over d\tau^2} =  
    \sum_{\beta,\gamma=0}^{3} {\partial {g}^{\beta\gamma} \over
    2 \partial x_\alpha} {d x_\beta \over d\tau} {d x_\gamma \over d\tau} 
 \eeq
which approximates to the non-relativistic equation of motion,
\beq
    {d^2 x_i \over d t^2} \approx -{\partial   {g}^{00} \over 2
    \partial x_i} c^2 \approx - \partial_i \Phi,
\eeq
where $d{x_0} = c dt \approx c d\tau$, and $g^{00} \approx -(1-2 \Phi/c^2)$.

The vector field is more fundamental than the scalar field
in TeVeS-like theories.  Any time-like vector field with four components can be approximated as 
\ba
A_\alpha &\approx& - e^{-\phi_{DE}+\Phi} (1 , 0, 0, 0) \\
g^{\alpha\beta} A_\alpha & \approx &  e^{-\phi_{DE}-\Phi} (1, 0, 0, 0),
\ea  
where we avoid the ambiguious notation of the upper index for the vector field, and
we omit the $c^2$ factor for more contact notations.
Here $\phi_{DE}$ is a scalar field, describing the modulus of the vector field $A$ 
\beq
-A^{2} \equiv -g^{\alpha\beta}A_\alpha A_\beta \equiv e^{-2\phi_{DE}}. 
\eeq
So the scalar field can be described by the physical metric $g^{\alpha\beta}$ and vector field $A_\beta$ alone.
The original proposal of Bekenstein contains
two metrics; the other metric (called Einstein metric $\tilde{g}$, 
where notations of tildes are opposite of Bekenstein) is fully 
described by the relation 
\beq
\tilde{g}_{\mu\nu} - A_\mu A_\nu = e^{2\phi_{DE}} {\tilde g}_{\mu\nu} - A_\mu A_\nu (2 - e^{4\phi_{DE}} ), 
\eeq  
The work of \cite{ZFS} shows that the TeVeS theory is 
equally described by a single physical metric $g^{\mu\nu}$, 
whose geodesics particles and light will follow.  All the effects 
of the vector field potential $A$ can be lumped together as 
a sophisticated Dark Energy like term.  E.g., 
the vector field contributes an E\&M-like Lagrange density 
$F_{\alpha\beta}F^{\alpha\beta}$, 
where from the covariant derivative of the vector potential $A$, 
one can form the Maxwell tensor field $F_{\alpha\beta}$
\beq
\tilde{g}_{\alpha\gamma} F^\gamma_\beta  
= F_{\alpha\beta}=\nabla_\alpha A_\beta - \nabla_\beta A_\alpha, 
\eeq
similar to the electric and magnetic field in electromagnetism.  This makes TeVeS in 
the similar framework as dark energy theories.  
From this perspective, one has not modified gravity.  One simply 
have a sophisticated energy term to replace the cosmological constant in the GR framework.

\subsection{Uneven Dark Energy fluid as Cosmological constant and as galaxy Dark Matter}

While Bekestein's original TeVeS Lagrangian is able to yield 
reasonable fits to CMB \cite{Sko}, there is an intrinsic discontinuity in its 
original proposal.  Zhao \& Famaey \cite{ZF} proposed to modify TeVeS Lagrangian 
to ensure a smooth transition between galaxies and cosmology.

\begin{figure}
\resizebox{13cm}{!}{\includegraphics{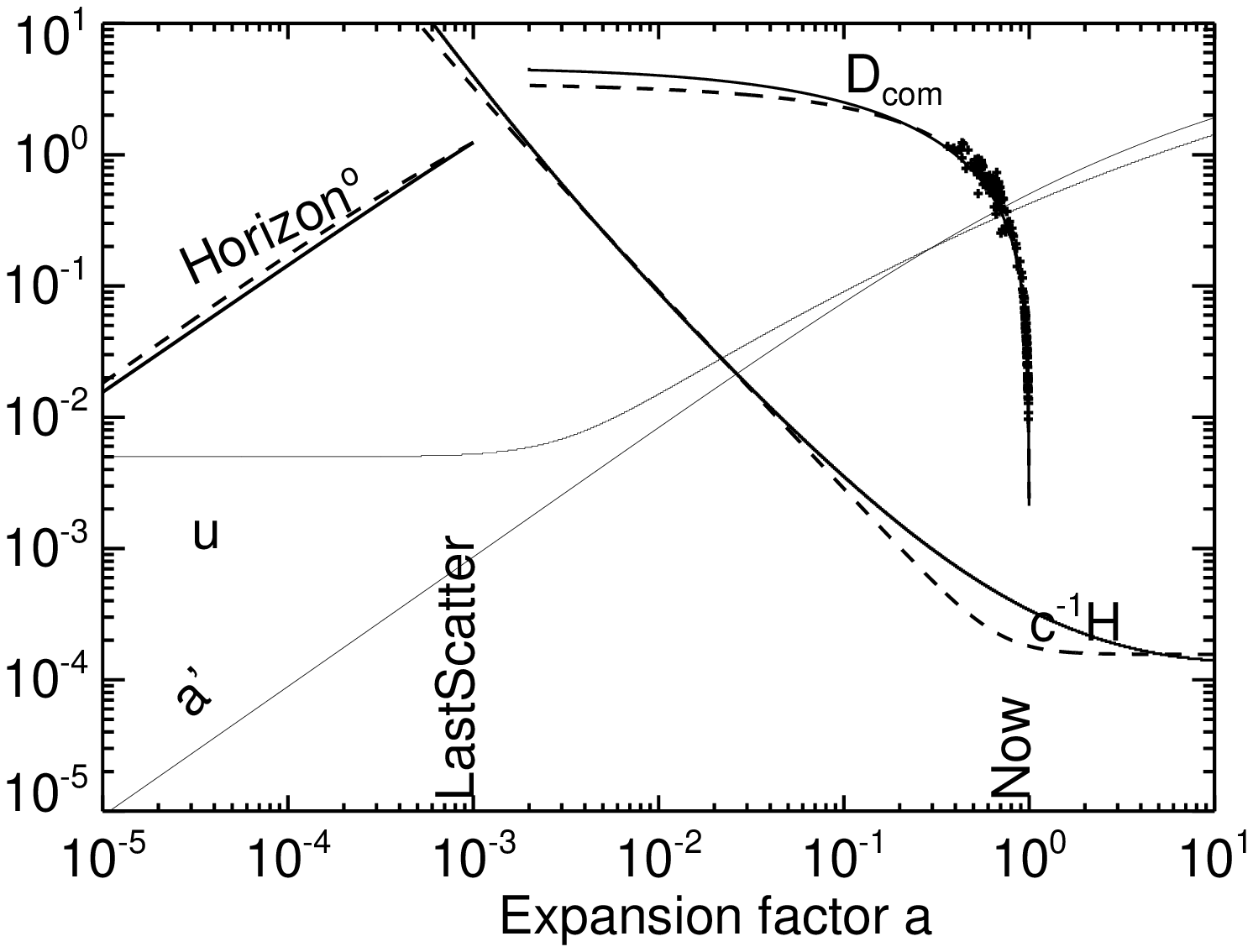}}
\vskip -1cm
\resizebox{13cm}{!}{\includegraphics{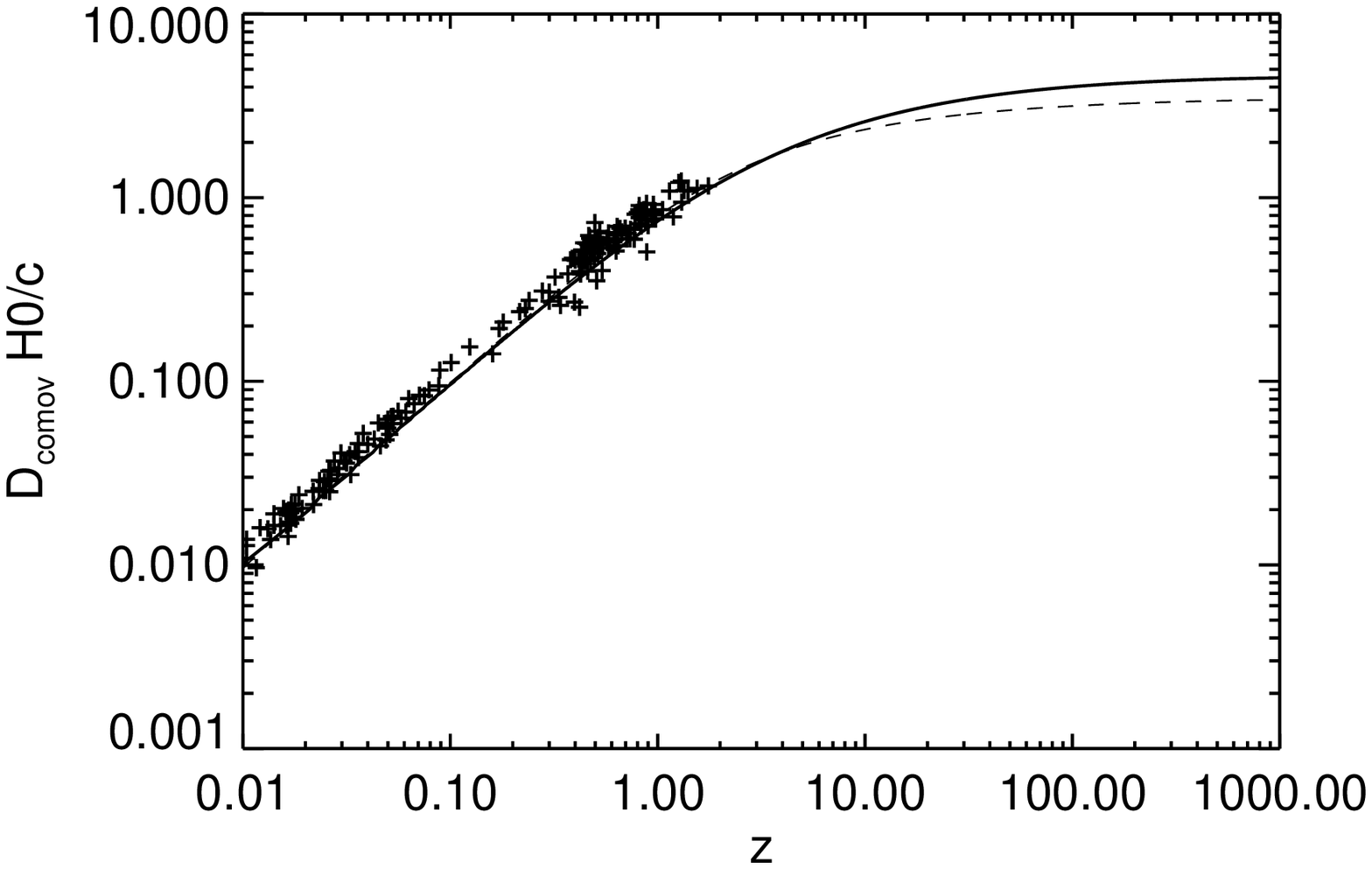}}
\caption{{\it Upper panel}: 
compares $\Lambda$CDM (dashed) with a TeVeS flat cosmologies without $\Lambda$ (solid) assuming zero mass for neutrinos and a $\mu$-function with $\alpha=0$.
Shown are the co-moving distance $D_{com}$ 
vs. the physical scale factor $a$ in log-log diagram overplotted with 
SNeIa data (small symbols) up to redshift 2.  
Likewise shows the horizon, the Hubble parameter $H$ in units of (${\rm Mpc}^{-1}c$) in two theories.   The evolution of the Dark Energy scalar field $\phi_{DE}$ and $\mu$  
can be inferred from (thin solid lines) $a' \equiv a \exp(\phi_{DE})$ and $u \equiv \mu^{-1}$.
{\it Lower panel}: Shows an enlarged view of the TeVeS fits to co-moving distance 
to the SNe data points.} \label{eureka}
\end{figure}

In the ZF proposal, the total action is that of the matter action $S_m$ 
plus Einstein-Hilbert action $S_{EH}$ 
plus the "cosmological constant"-like action for the vector field $A_\alpha$
\beq
S = S_m + S_{EH} + \int d^4x \sqrt{-{\tilde g}} {\Lambda \over 8\pi G}
\qquad {\Lambda \over 8\pi G} \equiv  
\left[ \int_0^\pounds {\mu_s d\pounds \over 8 \pi G_1} 
+ {1 \over 16 \pi G_2} F_\alpha^\beta F^\alpha_\beta  \right], 
\eeq
where the cosmological "constant" $\Lambda$ 
is replaced with {\it uneven} Dark Energy fields
\beq \label{pounds}
\mu_s \equiv  { f \over 1 - \alpha f}, ~f \equiv {\sqrt{\pounds} \over a_1}, ~
\pounds \equiv \left({\tilde g}_{\mu\nu} - A_{\mu}A_{\nu}\right)
{\nabla^\mu \phi_{DE} \nabla^\nu \phi_{DE} } ,
\eeq
where $\phi_{DE} = -{1 \over 2} \ln(- g^{\mu\nu} A_{\mu}A_{\nu} ) = {1 \over 2} \ln \sqrt{-{\tilde g} \over -g }$ is a scalar field which 
depends on the physical metric and the vector. 
Let the parameter $\alpha=0$ and adjust the constant parameters $a_1$, $G_1$, $G_2$, 
the model is able to fit approximately 
(cf. Fig.~\ref{eureka}) the late time acceleration 
from SNe without explicitly introducing a cosmological constant, 
and can explain the horizon scale angular size 
at recombination without explicitly introducing Dark Matter.  
This uneven DE fluid also satisfies the BBN constraints 
at $z \sim 10^9$, and the solar system constraints (see also \cite{Q2C} for details).  

This Lagrangian can also fit galaxy rotation curves at the present epoch without dark matter.  To see this, one can take variations of the action with respect to $A_\alpha$ and 
the metric $g_{\alpha\beta}$ respectively.  We get the vector field equation of motion
for this theory, and the Einstein equation for the dynamics of the metric tensor respectively.  The latter equation has the form
\bey
{G_{\alpha\beta} \over 8\pi G} &=& 
T^{K}_{\alpha\beta} 
+ T_{\alpha\beta}, \\
\eey
where the left-hand side is proportional to 
the Einstein tensor 
$G_{\mu\nu}  \equiv R_{\mu\nu}-{R \over 2} g_{\mu\nu}$ 
and on rhs the 1st term is the stress-energy tensor of known matter, 
the 2nd term is the stress-energy tensor for the vector field 
${T}_{\alpha\beta}$, which is a non-linear function of derivatives 
of the field $A_{\beta)}$.  
Near a galaxy, $G_{00}= 2 \grad\grad \Phi$.
Note that {\it the vector field stress tensor 
creates the mirage of additional matter}.

\subsection{TeVeS scalar field as effective dark matter}

In TeVeS, the galaxy potential $\Phi$ comes from two parts,
\beq
\Phi = \Phi_{K} + \phi_{DE}
\eeq
where the known Newtonian gravitational potential $\Phi_{K}({\bf x})$ 
of known matter of density $\rho_{K}({\bf x})$ satisfies
\beq
\grad \cdot \grad \Phi_{K} = 4 \pi G \rho_{K}
\eeq 
and the added scalar field satisfies
\beq
\grad \left[ \mu_s  \grad \phi_{DE} \right] = 4 \pi G \rho_{K}.
\eeq
Our Lagrangian free-function (eq.~\ref{pounds}) 
corresponds to the $\mu$-function proposed by \cite{ASZF} that  
\begin{equation}
        \mu_s = {f \over 1 - \alpha f}, \qquad 
f= \left|{\grad\phi_{DE} \over a_0} \right|. 
\end{equation}
This one-parameter $\alpha$ family of functions recovers
Bekenstein's\cite{Bek} toy model and the simple model of 
Zhao \& Famaey\cite{ZF} if setting $\alpha=0$ and $\alpha=1$ respectively.

Note if $\alpha=0$
\beq
\mu_s= \left|{\grad\phi_{DE} \over a_0} \right|, \qquad |\grad\phi_{DE}| = \sqrt{g_{K}a_0} 
\eeq
This way we recover the observed DM effects in eq.~(\ref{gdm}) 
and the classical MOND effect, i.e., 
the gravity $|\grad \Phi|$ drops as $\sqrt{GMa_0}/r$ far away from a point mass $M$.
Indeed this $\mu$ functions is able to fit rotation curves of 
faint and bright spiral galaxies approximately (cf. Fig.~\ref{vclens}),
although models with $\alpha=1$ fit better. \cite{FB06}

The picture to keep in mind is that 
the scalar field replaces the usual role of the potential of the Dark Matter.
The vector field $A$ is fully specified once $\phi_{DE}$ and $\Phi$ are given.

\subsection{Different interpolating functions: MOND vs TeVeS}

The gravitational potential in the classical MOND theory satisfies a modified Poisson's  equation,
\begin{equation}
\nabla[ \mu \nabla \Phi] = 4 \pi G \rho_{K}
\end{equation}
where the $\rho_{K}$ is the density of all known matter,
where $\mu$ is a function of total gravity.  This is 
different from TeVeS, where 
the total potential is the sum of Newtonian potential ($\Phi_N$) 
and a potential due to a scalar field ($\phi_{DE}$).
TeVeS $\mu_s$ is a function of the scalar field strength $g_s=|\nabla \phi_{DE}|$, and is derived from a free function in the action of the scalar field. 
In spherical symmetry, the two 
interpolation functions are related by 
\begin{equation}
\mu=\frac{\mu_s}{1+\mu_s}, \qquad  
g_{DE} = |\nabla \phi_{DE}| = {|\nabla \Phi_{K}| \over \mu_s} 
     = (1-\mu) \nabla \Phi = |\nabla \Phi - \nabla \Phi_{K}| 
\end{equation}
where $g_{DE}$ is the effective Dark Matter gravity due to a non-uniform Dark Energy (DE) field.

The standard MOND interpolating function $\mu(x)=\frac{x}{\sqrt{1+x^2}}$
is often used in fitting rotation curves.  But Zhao \& Famaey \cite{ZF}
argued that this function has undesirable features in TeVeS.  
For spherical systems our Lagrangian corresponds to a MOND function 
\begin{equation}
\mu(x)=\frac{2x}{1+(2-\alpha x)+\sqrt{(1-\alpha x)^2+4x}}, \qquad x=\left|{\grad\Phi \over a_0} \right|.
\end{equation}

\section{Light Bending in Slightly Curved Space Time}

Light rays trace the null geodesics of the space time metric.
Lensing, or the trajectories of light rays in general, are
uniquely specified once the metric is given.  In this sense light
bending works {\it exactly} the same way in any relativistic theory as
in GR.

Near a quasi-static system like a galaxy, the physical space-time
is only slightly curved.  Consider lensing by the galactic potential
$\Phi({\boldmath r})$.  A light ray moving with a constant
speed $c$ inside follows the null geodesics 
$d{t} = \sqrt{-\frac{\tg^{11}}{\tg^{00}}} dl$. 
An observed light ray travels a proper
distance $l_{os}=l_{ls}+l_{ol}$ from a source to the lens and then
to an observer. Hence it arrives after a time interval (seen by an
observer at rest with respect to the lens)
$
       \int d{t} = \int_{0}^{l_{os}} {dl \over c} - \int_{0}^{l_{os}}
       {2\Phi({\bf r}) \over c^2} {dl \over c}
$
containing a geometric term and a Shapiro time
delay term due to the $\Phi$ potential of a galaxy. 

In fact, gravitational lensing in TeVeS recovers many familiar
results of Einstein gravity in (non-)spherical
geometries.  Especially an observer at redshift $z=0$ sees a
delay $\Delta t_{\rm obs}$
in the light arrival time due to a thin deflector at $z=z_l$
 \beq\label{tdr}
    {c\Delta t_{\rm obs}({\bf R}) \over (1+z_l)}  \approx
    {D_{s} \over 2D_{l}D_{ls}}
    \left({\bf R}- {\bf R}_s \right)^2
    - \int_{-\infty}^{\infty} \!\!\! dl {2\Phi({\bf R},l) \over c^2},
 \eeq
as in GR for a weak-field thin lens, $\Phi/ c^2 \ll 1$. A light
ray penetrates the lens with a nearly straight line segment (within the thickness of the lens) 
with the 2-D coordinate, $\R=D_l \thetab$, perpendicular to the sky, where $D_l(z_l)=l_{ol}/(1+z_l)$
is the angular diameter distance of the lens at redshift $z_l$,
$D_s$ is the angular distances to the source, and $D_{ls}$ is the
angular distance from the lens to the source.
The usual lens equation can be obtained from the gradient of the
arrival time surface with respect to $\R$.  i.e.,
\ba
x- {D_l D_{ls} \over D_s} \alpha_x(x,y) = x_s  
&\qquad & \alpha_x  =  \int_{-\infty}^{\infty} \!\!\! dl {2 \partial_x \Phi(x,y,l) \over c^2},\\\nonumber
y- {D_l D_{ls} \over D_s} \alpha_y(x,y) = y_s  
&\qquad & \alpha_y  =  \int_{-\infty}^{\infty} \!\!\! dl {2 \partial_y \Phi(x,y,l) \over c^2},
\ea
and the convergence $\kappa$ is related to the deflection $(\alpha_x,\alpha_y)$ by 
\beq
\kappa  = {D_l D_{ls} \over 2D_s }\left( \partial_x \alpha_x + \partial_y \alpha_y \right).
\eeq
Likewise we get standard formulae for the shear and amplification:
$\gamma_2 = D_l \partial_y \alpha_x$ and $\gamma_1 = {D_l \over 2}\left( \partial_x \alpha_x - \partial_y \alpha_y \right)$ and for the amplification 
$A^{-1} = (1-\kappa)^2 - \gamma_1^2 -\gamma_2^2$. 

\section{Differences in lensing by uneven DE fluid and by DM halo}

An interesting point is that in GR $\kappa$ is proportional to the projected surface density of known matter.  This is {\it not the case for a non-linear theory of gravity,
nor for GR but with an even DE fluid}.  We can express $\kappa$ into the 
critical density as follows,
\beq
\kappa = {\tilde{\Sigma}(x,y) \over \Sigma_{crit}}, 
\qquad \Sigma_{crit}^{-1} \equiv {4 \pi G D_l D_{ls} \over D_s c^2 }, 
\eeq
where we define an effective projected density as follows,
\beq
\tilde{\Sigma}(x,y) \equiv  
\int_{-\infty}^{\infty} \!\!\! dl \tilde{\rho}(x,y,l),
\eeq 
note the integrand is NOT the true matter volume density at (x,y,l), 
rather
\beq
\tilde{\rho}(x,y,l) \equiv {\nabla^2 \Phi(x,y,l) \over 4 \pi G}
= \rho_{K} +  \rho_{DE} > \rho_{K} 
\eeq
because $\Phi$ is the addition of two fields, 
and we have a non-uniform Dark Energy (DE) fluid from the $\phi_{DE}$ field,
\beq
\rho_{DE} ={\nabla^2 \phi_{DE}(x,y,l) \over 4 \pi G} .
\eeq
The DE fluid tracks the known matter $\rho_{K}$, because 
the TeVeS $\phi_{DE}$ field is determined by non-linearly with $\rho_{K}$    
\beq
\rho_{K} = {\nabla \left[ \mu_s \grad \phi_{DE}(x,y,l) \right] \over 4 \pi G}. 
\eeq

There are some important differences between lensing in TeVeS and in GR
and between lensing a DE fluid and real DM halo: 
the potential is different.  To demonstrate this explicitly, 
let's consider a special non-spherical case, 
e.g., a Kuzmin disk lens.  Here one can solve the TeVeS Poisson equations
analytically. 
Consider an edge-on razor-thin disk lens of the Kuzmin profile of
a typical length $b$.  In TeVeS theory with a $\mu_s = |\grad \phi_{DE}|/a_0$,  
the Kuzmin disk would acquire a potential 
\beq
\Phi(x,y,z)=\Phi_K + \phi_{DE} = -{GM \over r_1} + \sqrt{GMa_0}\ln r_1, 
\qquad r_1 \equiv \sqrt{(b+|y|)^2+x^2+z^2},
\eeq
where the effective halo $\phi_{DE}(x,y,z)$ is non-spherical; its gradient has 
a sudden jump across the plane $y=\pm 0$, meaning that there is   
a razor thin layer of Dark Energy fluid.  
The effective halo would yield a non-zero non-axisymmetric convergence 
\beq
\kappa_{DE}(x,y) = {\pi \sqrt{GMa_0} \over c^2} {D_l D_{ls}/D_s \over \sqrt{(b+|y|)^2+x^2} }.
\eeq
In GR an edge-on disk without dark halo would have zero convergence.
We could add a spherical halo of real Dark Matter 
\beq
\phi_{DM}(x,y,z) = \sqrt{GMa_0} \ln \sqrt{b^2+y^2+x^2+z^2},
\eeq
centered on the origin $(x,y,z)=(0,0,0)$
such that the GR model produces identical potential 
$\phi_{DM}(x,0,z)=\phi_{DE}(x,0,z)$, hence 
identical rotation curve in the equator, as the TeVeS model.  The corresponding axisymmetric convergence
\beq
\kappa_{DM}(x,y) = {\pi \sqrt{GMa_0} \over c^2} {D_l D_{ls}/D_s \over \sqrt{b^2+y^2+x^2}},
\eeq
which is slightly {\it bigger} than that of 
the TeVeS $\kappa_{DE}$; e.g., for a line of sight with an impact
parameter $(x,y)=(0,b)$, we find $\kappa_{DE} = \kappa_{DM}/\sqrt{2} = {\pi \sqrt{GMa_0} \over  c^2} {D_l D_{ls} \over 2b D_s}$.  Note that the lensing time delay 
between a pair of images satisfies the scaling 
\beq
H_0 \Delta t_{\rm obs} \propto 1-\kappa,
\eeq
so the smaller convergence in TeVeS could predict
a larger $H_0$ to fit the same time delay data than in CDM model.  
Hence an uneven DE fluid offers a new way to bring
consistency of $H_0 \sim 70$km/s/Mpc from Hubble Key project and $H_0 \sim 50$km/s/Mpc from CDM fits to the lensing time delay measurements.\cite{ZQ}  

The vertical force $\partial_y \phi_{DE}$ and the deflection angle $\alpha_y$ along the $x=0$ line of sight are also {\it bigger} in real spherical DM halos than in TeVeS effective halo.  These differences suggest that a combined lensing and kinematics
study of a lens could decide whether the DM effects are due to real dark matter 
or an effective halo of uneven dark energy fluid.  

\section{Lensing and Other Sanity Tests of TeVeS-like theories}

For lenses with almost co-linear double images in the CASTLES survey, 
Zhao, Bacon, Taylor, Horne\cite{ZBTH} conducted a detailed fit using 
spherical point or Hernquist profile lenses.  Cares have been taken 
in including the K-correction, the luminosity evolution with redshift, and
the possibility of significant gas and extinction from dust.
They applied two methods, using the image positions only, and using
the image amplifications.  They found that
the mass-to-$M_*$ ratios calculated using the two independent methods
closely agree, and most of the lenses are found to
have $M/M_*$ between 0.5 and 2.  This
shows that TeVeS is a sensible theory for doing gravitational lensing,
in agreement with statistical analysis of a larger sample of lenses.\cite{CZ}\cite{Chen}

\begin{figure}
\resizebox{13cm}{!}{\includegraphics{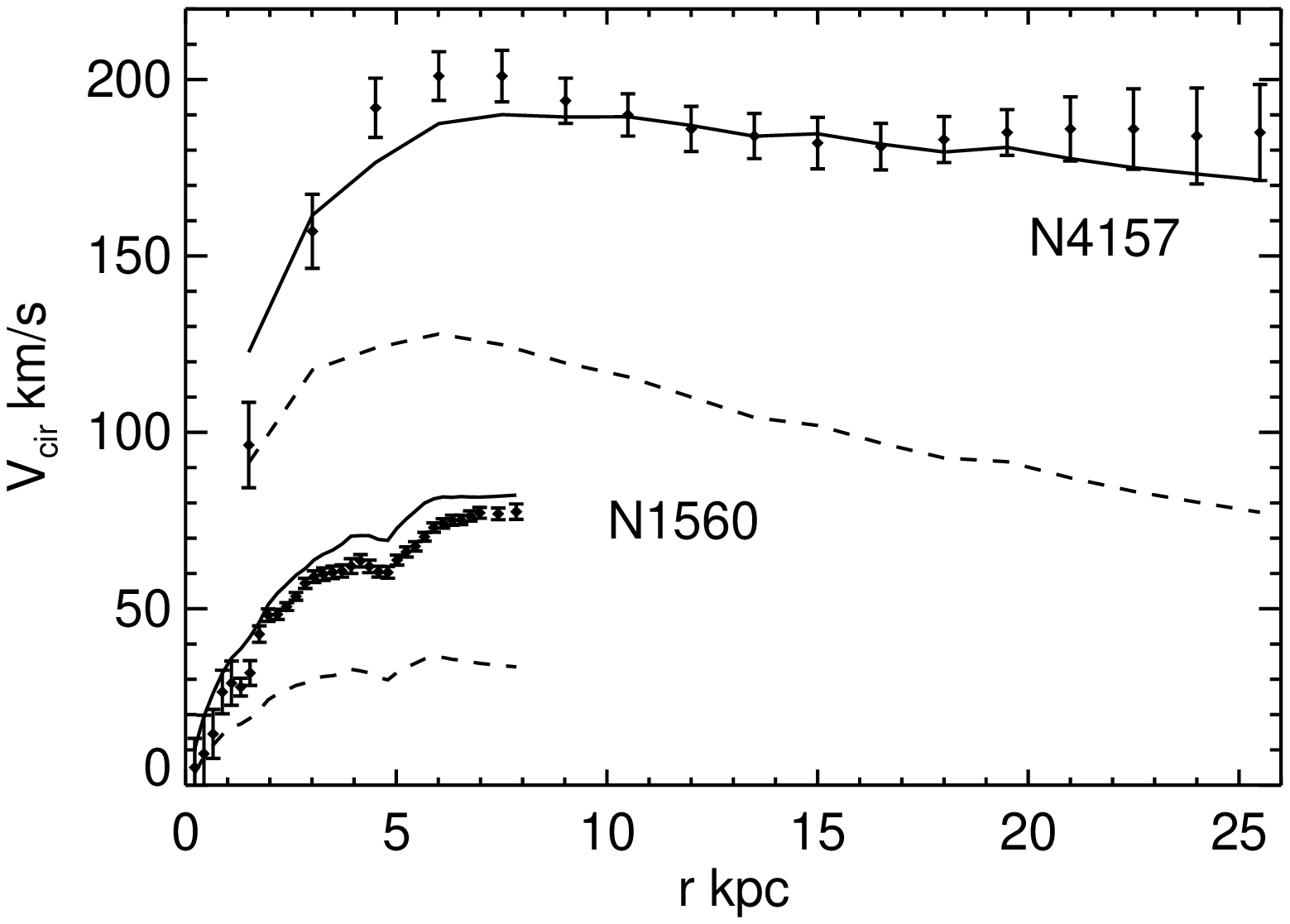}}
\vskip -1cm
\resizebox{13cm}{!}{\includegraphics{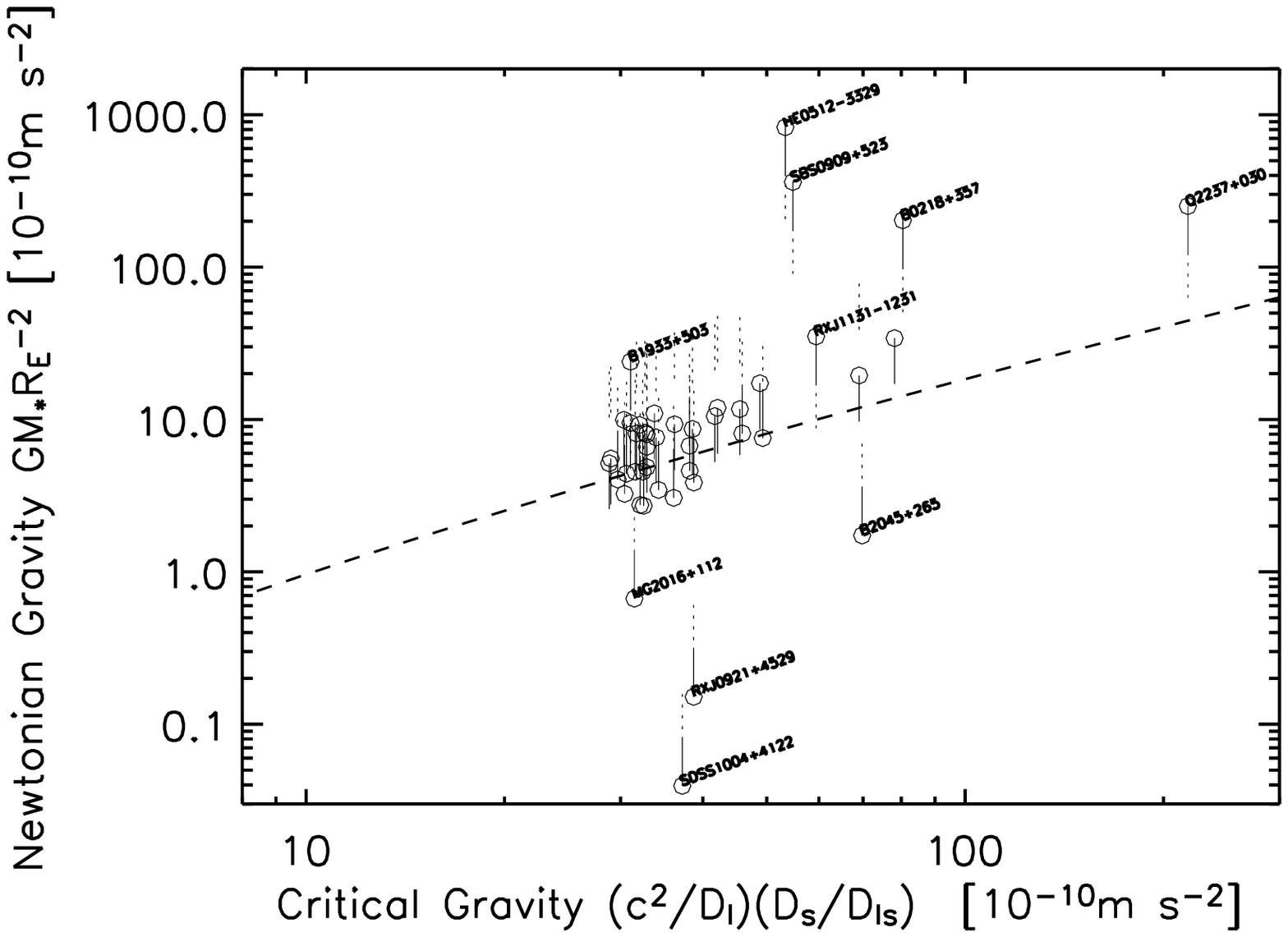}}
\caption{
{\it Upper panel}: Shows TeVeS baryon-only fits (solid)
to rotation curves of a gas-rich dwarf galaxy NGC1560 
($M_*/L_*=1.3$) and a gas-poor larger spiral galaxy NGC4157 ($M_*/L_*=0.6$),
and adopting $a_0=1.2\times 10^{-8}$, $\alpha=0$; 
the Newtonian $V_{cir}$ by baryons is also shown (dashed).  
{\it Lower panel}: Shows the scatter of two measurements of gravity near Einstein radii of about 50 CASTELS multi-imaged lenses.  The gravity due to stars (vertical axis) and the gravity observed (horizon axis) appear correlated around
a straight line for many lenses, as expected in TeVeS.  A few outliers are labeled, 
consistent with the recent analysis of Ferreras et al..
}\label{vclens}
\end{figure}

Nevertheless, I caution that there are several lenses (cf. Fig.~\ref{vclens}), typically in galaxy
clusters, which require extreme M/L, e.g.,\cite{ZBTH},\cite{Ferreras}.  
Outliers can occasionally be caused by photometry errors since the lens galaxy is barely resolved, and its total luminosity is subject to the uncertain subtraction of the much brighter quasar images around it.  On the modeling side, 
all previous models are spherical while the flattening and the external shear of 
real lenses are not taken into account.  
Also the cluster environment makes prediction highly uncertain: 
the cluster gas increases the total baryonic material in the lens, but 
the whole lens accelerates in the cluster, and this so-called external acceleration
\cite{Wu} decreases the MOND effect of the lens.  In general, our non-linear Poisson equation for $\phi_{DE}$ can be solved by 
adapting the numerical code of e.g., the Bologna \cite{Bologna} 
or the Paris\cite{Paris} group.

Less model-dependent one can ask if the gravity in stars are correlated 
with the image-splitting power of the lens; such a correlation is expected   
if TeVeS is correct.  Indeed Fig.~\ref{vclens} shows such a correlation.
The horizontal axis is proportional to the critical density 
$\Sigma_{crit} =c^2/(4\pi G D)$ needed to split images, where $D=D_lD_{ls}/D_s$.  
Most strong lenses are elliptical galaxies, and the projected surface density within the Einstein radii is much higher than ${a_0 \over 2 \pi G}$, hence the MOND effect 
is very mild, and the MOND effective halo is sub-dominant within the Einstein ring.
The vertical axis is proportional to the mean density of stars within the Einstein radii  
$\Sigma_*= M_*/(\pi R_E^2)$, assuming $M_*/L_*=4$ (circles); effects of raising/lowering $M_*/L_*$ by a factor of $2$ (solid) or a factor of $4$ (dotted) are shown by little vertical rods for each lens.  There are, however, quite a few lenses whose gravitation appear uncorrelated with its baryonic mass.  

For complicated lens geometry, one can also model lensing 
by starting with a reasonable guess for the 3D potential, 
and find the density by taking appropriate derivatives. E.g., 
Angus et al. \cite{ASZF} model the Bullet Cluster by a double-peaked 
potential and find that 
the lensing peaks of the Bullet Cluster could be explained by adding
neutrinos as part of the known density $\rho_{K}$
in a TeVeS-like modified gravity; there is also a tentative evidence
from galaxy rotation curves in the Ursa Major cluster\cite{GZF}.
The phase space density of
neutrinos at the lensing peaks requires at least 2eV mass for neutrinos 
in order not to violate exclusion principle for fermions.  

Sanity checks from the solar system to large scale have also been done 
in recent papers.  For example, TeVeS is found to be broadly consistent 
with galaxy dynamics of early-type galaxies and disk galaxies
(see references in \cite{AFTZCK} \cite{MS} \cite{SN}).  
and observations of vertical force, escape velocity and microlensing in the Milky Way (see references in \cite{NLZC} \cite{FBZ}). 
It is possible to build self-consistent triaxial elliptical galaxies
using the Schwarzschild method\cite{WWZ}.  It is possible to explain the rotation curve of tidal dwarf galaxies, which are hard to understand in CDM framework \cite{TDG}.  
Structures and CMB anisotropy can form from linear perturbations
(see references in \cite{Q2C}).  In general, a non-uniform dark energy fluid can mimic many effects of Dark Matter. \cite{Halle}

Nevertheless,
TeVeS-like theories are by no means a firmly established paradigm since
many comparisons of the theories with observations are still unknown.
While this is normal for a new theory, 
the Bullet Cluster and some outliers among gravitational lensing galaxies are worring.   
Also such theories face challenge to explain why globular clusters and 
dwarf galaxies of the same baryonic mass shows very different gravitational mass
\cite{Roche}\cite{ZT} unless the dark energy fluid is allowed to condense on sub-kpc scale.
In the process of understanding and falsifying TeVeS-like theories, 
we hope to learn to design more clever and robust emulators for dark matter effects.  It is worth stressing that a common goal of 
both the standard approach and alternative approach
is to understand the detailed physics of the vacuum energy.  
The (scalar or vector) fields in the vacuum 
might ultimately hold the answers to both DE and DM 
mysteries and the answers to 
many fundamental questions in particle physics.
 
\section*{Acknowledgements}

It is difficult to write a review and not to be outdated 
in less than a year in this rapidly developing field,
where ideas are constantly being falsified, and 
remerged with other ideas for synergy.  
I apologize for my incomplete survey of the vast literatures involved.  
The aim of this review is partially to stimulate better theories to emerge.  
I thank Benoit Famaey, Sean Carroll, Eugene Lim, Alan Kostelecky, 
Ted Jacobson, Jacob Bekenstein, Constantinous Skordis, David Mota, 
Pedro Ferriera, Tom Zlosnik, Glenn Starkman, 
Andy Taylor, Luca Ciotti, Carlo Nipoti, Francoise Combes, J-P Bruneton, 
Daming Chen, Baojiu Li, Garry Angus, Martin Feix and many other collegues
for stimulating discussions on cosmology, dynamics and lensing.
HSZ acknowledges partial support of PPARC Advanced Fellowship
and National Natural Science Foundation of China (NSFC under grant No. 10428308).

\end{document}